\documentclass[runningheads]{llncs}
\usepackage{graphicx}
\usepackage{verbatim}
\usepackage{array}
\newcolumntype{P}[1]{>{\centering\arraybackslash}p{#1}}
\usepackage{multirow}
\usepackage{subcaption}
\begin{document}
\title{Impact of Experiencing Misrecognition by Teachable Agents on Learning and Rapport}
\titlerunning{Impact of experiencing misrecognition by teachable agents}
\author{Yuya Asano \inst{1} \and
Diane Litman \inst{1} \and
Mingzhi Yu \inst{1} \and
Nikki Lobczowski \inst{2} \and
Timothy Nokes-Malach \inst{1} \and
Adriana Kovashka \inst{1} \and
Erin Walker \inst{1}
}
\authorrunning{Y. Asano et al.}
\institute{University of Pittsburgh, Pennsylvania, USA \\
\email{\{yua17, dlitman, miy39, nokes, aik85, eawalker\}@pitt.edu}
\and
McGill University, Quebec, Canada \\
\email{nikki.lobczowski@mcgill.ca}
}
\maketitle              % typeset the header of the contribution
\begin{abstract}
While speech-enabled teachable agents have some advantages over typing-based ones, they are vulnerable to errors stemming from misrecognition by automatic speech recognition (ASR). These errors may %or may not 
propagate, resulting in unexpected changes in the flow of conversation. We analyzed how such changes are linked with learning gains and learners' rapport with the agents. 
%Our results show they are not related to learning gains but that learners felt less rapport when agents said something generic given that they should have said something more specific to the problems they were taught.
Our results show they are not related to learning gains or rapport, regardless of the types of responses the agents should have returned given the correct input from learners without ASR errors.
We also discuss the implications for optimal error-recovery policies for teachable agents that can be drawn from these findings.

\end{abstract}

\section{Introduction and Related Work}

Students benefit from teaching others more than being tutored \cite{annis1983processes}. This effect of learning by teaching also holds when they teach a virtual agent \cite{leelawong2008designing} or an embodied robot \cite{lubold2018producing} (called a teachable agent or robot). Although interaction with teachable agents can be done through typing \cite{lee2021curiosity} or speech \cite{asano-etal-2022-comparison}, the literature suggests speech-based interaction is powerful in tutorial dialogues in general because speech %contains prosodic and meta-communicative information \cite{pon2004advantages} and 
enables learners to complete tasks faster than typing due to ease of production \cite{dmello2010toward}. However, speech-based teachable agents are susceptible to misrecognition made by automatic speech recognition (ASR) when converting speech input to text. It may change the flow of the dialogue between students and agents and thus affect students' learning and perception of the agents.

\begin{figure}[t]
\centering
\includegraphics[width=0.82\textwidth]{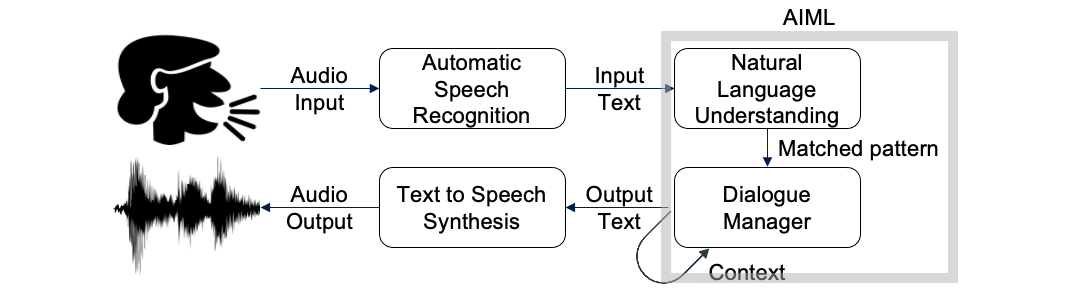}
\caption{The structure of our dialogue system. When it receives audio input, it converts audio to text. Then, Artificial Intelligence Markup Language (AIML) finds a pattern in the input text (NLU), selects an output (Dialogue Manager), and remembers it as a context. Finally, the system converts the text to speech.} \label{dialogue}
\end{figure}

Past work has explored how misrecognition in different stages of speech-enabled tutorial dialogue systems (Fig. \ref{dialogue} shows the structure of our system) is related to students' learning gain and evaluation of the systems. D'Mello et al. \cite{dmello2010toward} have found word error rates of ASR were not associated with students' learning gain but were related to their satisfaction with systems. Litman and Forbes-Riley \cite{litman2005speech} have found no correlations between errors made by a natural language understanding (NLU) module and students' learning. However, errors in earlier stages may not propagate to the outputs of a system. For example, mistakes in singular versus plural forms or ``a'' versus ``the'' made by ASR have little effect on NLU. Moreover, even if NLU fails to map ``massive'' to ``big'' because ``massive'' is out of vocabulary, the output may not change if it does not care whether a user says ``big''. Thus, errors in ASR or NLU may not represent misrecognition felt by end users. In this paper, we focus on mistakes that are retained until the final outputs of systems. 
Also, Dzikovska et al. \cite{dzikovska-etal-2010-impact} have shown that the frequency of a system not understanding user inputs and thus replying with a neutral response is negatively correlated with user satisfaction but not with learning gain. %Still, they do not 
We instead distinguish the case where it is fine for dialogue systems to return a neutral response because learners say something irrelevant from the case where the systems should respond with something more specific. 

We fill a %missing piece of the 
literature gap on the effect of errors by dialogue systems on learners. % in two ways. 
First, we extend the work on tutoring systems that appear more knowledgeable than learners to teachable agents that are at most the same level. %as learners. 
Second, we analyze errors observable by learners, instead of %the ones that happen internally in systems.
errors internal to systems. 
%Specifically, we give raw ASR inputs and ``true'' inputs from users to an agent %teachable NAO robot %named Emma 
%and compare its responses. 
We define \textit{dialogue misrecognition} as the situation where an agent responds differently to raw ASR inputs and ``true'' inputs. Finally, we evaluate an agent with students' sense of rapport with it rather than user satisfaction. Rapport is a predictor of learning for both human and agent tutees \cite{lubold2018producing} and leaves a positive impression that helps agents establish a long-term relationship with learners \cite{gulz2011extending}. Thus, it is likely a more direct metric than satisfaction with the effectiveness of human-agent collaborations. We have found dialogue misrecognition is not linked with learning or rapport. Our results are in line with the literature but contribute to it by measuring only misrecognition that impacts the flow of conversation and by using an outcome more suitable for human-agent collaborations.

\section{Method}
\subsection{Dataset}

We used the dataset from an experiment where 40 undergraduate students (35 female, 5 male; 17 White, 13 Asian, 5 Black, 1 Latino, 4 unknown; mean age $= 19.64$, $SD = 1.25$) in a US city taught ratio word problems to a robot named Emma using spoken dialogue for 30 minutes. %(see Fig. \ref{emma}) 
The study was over Zoom due to COVID-19. To talk to Emma, they pushed and held a button on a web application. She is designed to follow up on student explanations with questions or her own explanations. For example, she says \textit{``So I multiply because I have three times as many people?''} after a student tells \textit{``You have to multiply cups of seltzer by three.''} She also guides students to a correct solution when they are unsure about the answer. For instance, if a student's utterance contains \textit{``I don't know''}, she suggests the next step, saying \textit{``Me either. Maybe I start by dividing?''}. We had two experimental conditions: students teaching in pairs (n=28) or alone (n=12). Our design was quasi-experimental; if only one of the students showed up out of a pair, we had them teach Emma alone. Students in pairs taught her collaboratively by alternating between discussing the problems with their partners and talking to her. We excluded one pair from our analysis because one of the students did not speak to Emma during the session.

%\begin{figure}
%\centering
%\includegraphics[width=0.75\textwidth]%{images/emma_screen.png}
%\caption{Screenshot of two students teaching %Emma.} \label{emma}
%\end{figure}

The students individually took a pre-test before the session and a post-test and a survey after that to assess learning and rapport with Emma.  We prepared two versions of pre- and post-tests each of which had 13 ratio problems similar to what they saw while teaching her. We removed five isomorphic problems from each test that proved to have different difficulties between versions, leaving eight problems for analysis \cite{steele2022takes}. 
On average, students scored 5.58 ($SD = 1.90$) in the pre-test and 6.82 ($SD = 1.35$) in the post-test. Rapport was measured on a six-point Likert scale devised by \cite{lubold2018producing} (mean $= 4.49$, $SD = .679$) and represented the average of items asking about mutual positivity, attention, and coordination between the student and the robot \cite{lubold2018producing,asano-etal-2022-comparison}.

\subsection{Quantification of dialogue misrecognition}

We simulated Emma's responses to ``true'' inputs as follows. First, we manually transcribed the students' utterances directed to her. Let $U^H_t$ be the human transcribed utterance of a student directed to Emma in turn $t$ and $U^{ASR}_t$ be the utterance transcribed by ASR. Next, we sent $U^H_t$ to Emma to get a simulated ``true'' response $R^H_t$. She selects $R^H_t$ from a set of pre-authored responses written in Artificial Intelligence Markup Language (AIML), based on patterns in $U^H_t$ and context $C_t$. We set $C_1$ to none and $C_{t+1}$ to $R^{ASR}_t$, the response to $U^{ASR}_t$, because $U^H_{t+1}$ is a reply to $R^{ASR}_t$. If there is no matching pattern in $U^H_t$ given $C_t$, Emma returns an utterance randomly selected from a set of \textit{generic} responses $G$ that can make sense in any contexts such as \textit{``I think I get it. What do I do next?''}. The size of $G$ was 28. The responses in $G$ do not change Emma's state, meaning that, unlike non-generic responses, they do not let students move to the next step of the problem. Finally, we calculated the proportions of the turns in which $R^H_t$ differ from $R^{ASR}_t$ to the total number of turns for each student (i.e. $P(R^H_t \neq R^{ASR}_t)$, the proportion of \textbf{Overall} dialogue misrecognition). We did not use raw numbers of errors to normalize the differences in the number of turns for each student. In the case of students in pairs, we did not use the turns where their partner spoke to calculate the proportions but did use those turns to define $C_{t+1}$ and update Emma's dialogue manager. Note that we treated $R^H_t$ and $R^{ASR}_t$ as the same when both are from $G$, even if their surface forms are different.
An example of the calculation of the proportions is shown in Table \ref{example}.

\begin{comment}
\begin{table}[]
    \centering
    \caption{Example interaction between students in a pair and Emma. Her responses in italics come from the set of generic responses $G$. In this example, each student had one turn, and Emma misrecognized student A's input. Therefore, $P(R^H_t \neq R^{ASR}_t) = 1$ for student A and $P(R^H_t \neq R^{ASR}_t) = 0$ for student B.}
    \begin{tabular}{c|P{3.4cm}|P{2.6cm}|P{2.3cm}|c}
         Speaker & $U^H_t$ & $R^H_t$ & $R^{ASR}_t$ & Misrecognition \\ \hline
         Student A & Emma, in step zero, you're going to take the new amount and divide it by the original amount. So what is twelve divided by four to find the ratio? & So I have three times as many people. So maybe I multiply? Because I have three times as many people. & \textit{I'm not sure actually.} & $R^H_t \neq R^{ASR}_t$ \\ \hline
         Student B & Well, can you divide twelve by four? & \textit{I'm still learning this. I don't get it.} & \textit{Can you explain a little more?} & $R^H_t = R^{ASR}_t$
    \end{tabular}
    \label{example}
\end{table}
\end{comment}

\begin{table}[t]
    \centering
    \caption{Example interaction between students in a pair and Emma. Her responses in italics come from the set of generic responses $G$. In this example, each student had one turn, and Emma misrecognized student B's input. Therefore, $P(R^H_t \neq R^{ASR}_t) = 0$ for student A and $P(R^H_t \neq R^{ASR}_t) = 1$ for student B.}
    \begin{tabular}{l|l|p{7cm}|p{2.5cm}}
         Speaker & Label & Utterance & Results \\ \hline
         A & $U^{ASR}_1$ & what is the ratio between the volume of paint and the surface area & AIML pattern: no match\\ \cline{2-4}
         & $U^H_1$ & Emma, what is the ratio between the volume of paint and the surface area? & AIML pattern: no match\\ \cline{1-4}
         Emma & $R^{ASR}_1$ & \textit{I'm still learning this. I don't get it.} & \multirow{2}{2.5cm}{$R^{ASR}_1 = R^H_1$} \\ \cline{2-3}
         & $R^H_1$ & \textit{I'm not sure actually.} &  \\ \hline
         B & $U^{ASR}_2$ & the ratio between surface area and volume is 622 & AIML pattern: no match\\ \cline{2-4}
         & $U^H_2$ & The ratio between surface area and volume is 6 to 2. & AIML pattern: ``ratio ... 6 ... 2''\\ \cline{1-4}
         Emma & $R^{ASR}_2$ & \textit{Can you explain a little more?} & \multirow{2}{2.5cm}{$R^{ASR}_2 \neq R^H_2$} \\ \cline{2-3}
         & $R^{H}_2$ & 6 to 2 is the same ratio I used for step one. But the ratio of 1 to 3 seems an easier place to start? & 
    \end{tabular}
    \label{example}
\end{table}

We further categorized dialogue misrecognition into three cases:
\begin{itemize}
    \item \textbf{Prevented}: $P(R^H_t \notin G \land R^{ASR}_t \in G)$. This means the student could not move to the next step because Emma misrecognized their input.
    \item \textbf{Different}: $P(R^H_t \neq R^{ASR}_t \land R^H_t \notin G \land R^{ASR}_t \notin G)$. This often implies Emma suggested a different way to solve a problem from what the student said.
    \item \textbf{Proceeded}: $P(R^H_t \in G \land R^{ASR}_t \notin G)$. This represents the case where Emma went to the next step by accident due to misrecognition.
\end{itemize}

\section{Results and Discussion}

We examined how dialogue misrecognition is related to students' rapport with Emma and learning gain by running correlation analyses. We used Pearson's correlations for rapport and partial correlation for post-test scores controlled by pre-test scores because pre-test scores were positively correlated with post-test scores ($r=.443$, $p=.005$). Table \ref{all} summarizes our analysis of the 38 students from both conditions. \textbf{Overall} misrecognition is not correlated with either rapport ($r=-.016$, $p=.922$) or learning ($r=.246$, $p=.142$). Of the three types of dialogue misrecognition, \textbf{Different} was the highest. None of these types was significantly correlated with learning or rapport. However, \textbf{Prevented} is marginally negatively correlated with rapport ($r=-.319$, $p=.051$).

\begin{table}[t]
    \centering
    \caption{Descriptive statistics of the proportions of dialogue misrecognition and its correlations with rapport with Emma (Pearson's r, $df=36$) and post-test scores (partial correlation controlled by pre-test scores, $df=35$).  \textbf{Different} type was the most likely. No correlation was significant.}
    \begin{tabular}{c|c|c|c}
         Misrecognition types & Mean (SD) & Rapport (p-value) & Learning (p-value) \\ \hline
         \textbf{Overall} & .146 (.080) & -.016 (.922) & .246 (.142) \\ \hline
          \textbf{Prevented} & .026 (.031) & -.319 (.051) & .188 (.488) \\ \hline
          \textbf{Different} & .103 (.070) & .141 (.399) & .271 (.104) \\ \hline
          \textbf{Proceeded} & .017 (.025) & -.056 (.737) & -.131 (.440)
    \end{tabular}
    \label{all}
\end{table}

Our results provide another piece of evidence that misrecognition by teachable agents is not necessarily relevant to learning gains or learners' perception of agents. Furthermore, this implies that an optimal error-recovery strategy for teachable agents should prefer moving to the next step, assuming that learners give reasonable inputs, rather than expressing they do not understand the inputs. This may sound counterintuitive because it may deprive learners of opportunities to realize their misunderstanding. Still, this disadvantage may be canceled out by exposure to correct solutions and more problems because the correlation between learning and the proportion that Emma returns a non-generic response when she is supposed to return a generic one is not significant. This policy may also aid inclusion because it can avoid generic responses stemming from ASR's poor performance in accented speech and minoritized dialects.

One limitation of this study is that many participants were at the ceiling (6 students scored 100\% on the pre-test) and thus did not learn as part of the study, reducing our ability to examine correlations between dialogue misrecognition and learning. Another limitation is that our dialogue system is not a state-of-the-art end-to-end neural model. %is simple, and thus these findings might not generalize to dialogue systems with more complex interactions that inspire users to have higher expectations. 
We used the Web Speech API for speech recognition off the shelf, which yielded a .226 word error rate on average ($SD = .102$)\footnote{Word error rates were not correlated with rapport ($r=.196$, $p=.239$), learning ($\rho=.246$, $p=.142$), or overall dialogue misrecognition ($r=-.155$, $p=.354$).}, and performed only pattern matching to decide Emma's response. Yet, our dialogue misrecognition measures can be used for end-to-end models that do not have internal components such as NLU. Also, due to the small sample size ($n=38$), we lack statistical power and could not include demographic variables or the experimental conditions as covariates. This stopped us from analyzing the effect of witnessing dialogue misrecognition encountered by a partner.

We proposed new measures of dialogue misrecognition to explore how changes in a conversation flow caused by errors that propagate through a dialogue system are related to rapport with teachable agents and learning gain. Our results indicate these changes are not linked to learning or rapport. This implies we do not need a sophisticated dialogue system with little misrecognition for teachable agents and that an optimal error-recovery policy can be as simple as presuming inputs from learners are reasonable. Future research can test how this policy affects rapport, learning, and other outcomes such as engagement.

\section{Acknowledgments}
We would like to thank anonymous reviewers for their thoughtful comments on this paper. This work was supported by Grant No. 2024645 from the National Science Foundation, Grant No. 220020483 from the James S. McDonnell Foundation, and a University of Pittsburgh Learning Research and Development Center internal award.
\bibliographystyle{splncs04}
\bibliography{bibliography}
\end{document}